\documentclass[12pt]{iopart}

\usepackage{graphicx}
\usepackage{tabularx}
\usepackage{multirow}
\usepackage{hyperref}
\usepackage{amsmath}
\usepackage{mathrsfs}
\usepackage{amsfonts}

\begin{document}

\title[Transient absorption in LNT]{Long-lived, pulse-induced absorption in LiNb$_{1-x}$Ta$_x$O$_3$ solid solutions: the case of three intrinsic defect sites for electron localization with strong coupling}

\author{Niklas Dömer$^{1}$*, Julian Koelmann$^{1}$*, Mira Hesselink$^{1}$, Tobias Hehemann$^{1}$, Anton Pfannstiel$^{1}$, Felix Sauerwein$^{1,2}$, Laura Vittadello$^{1,2}$, Steffen Ganschow$^{3}$ and Mirco Imlau$^{1,2}$}

\address{$^{1}$ Department of Mathematics/Informatics/Physics, Osnabrueck University, 49076 Osnabrueck, Germany}
\address{$^{2}$ Research Center for Cellular Nanoanalytics, Osnabrueck (CellNanOs), Osnabrueck University, 49076 Osnabrueck, Germany}
\address{$^{3}$ Leibniz-Institut für Kristallzüchtung (IKZ) im Forschungsverbund Berlin e.V., Max-Born-Straße 2, 12489 Berlin, Germany}
\begin{indented}
\item[]
\end{indented}
\begin{indented}
\item[]*Authors contributed equally to this work
\end{indented}
\ead{mirco.imlau@uni-osnabrueck.de}
\vspace{10pt}
\begin{indented}
\item[]March 2024
\end{indented}

\begin{abstract}
Femto-/nanosecond pulse-induced, red and near-infrared absorption is studied in LiNb$_{1-x}$Ta$_{x}$O$_3$ (LNT) solid solutions with the goal to probe the intrinsic defect structure via the formation, transport and recombination of optically generated small bound electron polarons with strong coupling to the lattice. 
As a result, long-lived transients are uncovered for LNT which exceed lifetimes of LN and LT by a factor of up to 100 over the entire range of investigated compositions.
At the same time, the starting amplitude varies in the range of $\alpha_{\rm li}^0\approx10-100$\,m$^{-1}$ as a function of $x$ and exceed the ones of LN and LT by a factor of up to ten. The results are interpreted in the model of three-dimensional small polaron hopping transport considering the simultaneous presence of three different types of small bound polarons, in particular of small electron Nb$_{\rm Li}^{4+}$ and Ta$_{\rm Li}^{4+}$ antisite polarons, and of small electron Ta$_{\rm V}^{4+}$ interstitial polarons. 
We conclude that the differences between LNT, LN, and LT may point to model systems that consist of one (LN), two (LT) and three (LNT) intrinsic defect centers for electron localization. 
\end{abstract}

\maketitle

\section{Introduction}
Pulse-induced transient absorption is a powerful tool for the inspection of the intrinsic defect landscape of polar oxide materials, including the application-relevant ferroelectrics lithium niobate (LiNbO$_3$, LN)~\cite{weis-apa85} and lithium tantalate (LiTaO$_3$, LT)~\cite{Abrahams-jpcs67}. The absorption is attributed to optically generated small polarons, i.e. charge carriers that are self-localized at lattice sites by Coulomb interaction and electron-phonon coupling~\cite{Emin2013}. In LN and LT single crystals grown from the congruent melt (c-LN and c-LT in what follows), the phenomenon is particularly pronounced with broad-band absorption features (FWHM $\sim 1$\,eV) in the near-infrared (NIR) spectral region~\cite{Merschjann-jpcm09} because the localization takes place at the intrinsic defect centers, at the Nb$_{\rm Li}^{5+}$ antisites (LN) (E$_{\rm p}^{\rm opt}=1.6$\,eV)~\cite{Schirmer2009}, at the Ta$_{\rm Li}^{5+}$ antisites (E$_{\rm p}^{\rm opt}=2.0$\,eV)~\cite{Krampf2021} and interstitial Ta$_{\rm V}^{5+}$:V$_{\rm Li}$ defect pairs (E$_{\rm p}^{\rm opt}=1.5$\,eV)~\cite{pfannstiel-jpcm24} (LT), respectively. The time-resolved inspection of the absorption kinetics enables insight to the small-polaron formation dynamics on the sub-ps time scale~\cite{Freytag2018}, the transport of small polarons via 3D hopping within the polar crystal structure on the sub-microsecond time scale~\cite{Guilbert-jpcm18} including intermediate trapping at intrinsic and extrinsic defect sites~\cite{Messerschmidt-ome19} and the death of small polarons by electron-hole-recombination after about a few micro-/milliseconds~\cite{berben-jap00}. The proper analysis reveals the basis for the control of the intrinsic landscape by doping with extrinsic elements that foster application relevant properties. Prominent examples are the incorporation of Magnesium to suppress the optical damage~\cite{Volk-josab94}, of Titanium for the production of waveguides~\cite{Bazzan-apr15}, of Erbium for the production of light sources~\cite{Brinkmann-el91} and of Iron for photorefractive applications such as phase-conjugated mirrors~\cite{Günter-b07}. In most of the cases, the incorporation of the dopant ions takes places on lithium sites~\cite{Volk-b09, Fontana-apr15}, thus minimizing the number density of intrinsic antisites for small polaron localization with a significant impact on the near-infrared absorption properties~\cite{Conradi-pssrrl08}. 
The analysis of the transient absorption also provides a comprehensive insight to the fundamental relation of the crystal physical properties with the intrinsic defect structure that includes the electronic transport phenomena in congruent, (near-) stoichiometric and doped single crystals of LN and LT, but also the variety of optical nonlinearities 
related to the formation of small polarons with strong coupling~\cite{Imlau2015}.

We here present our study on pulse-induced, near-infrared absorption in the model system lithium niobate tanatalate, LiNb$_{1-x}$Ta$_x$O$_3$ (LNT, $0\leq x\leq 1)$. LNT is a promising material due to the possibility of adjusting the structural, (photo-)electrical, mechanical, ionic and (non-)linear optical properties via the composition within the parameter space of its edge compositions lithium niobate (LN, $x=0$) and lithium tantalate (LT, $x=1$). 
As a first attempt LNT can be regarded as a mixture of LN and LT. In this respect, it is expected that the intrinsic defect structure of LNT contains Nb$_{\rm Li}$ antisites, Ta$_{\rm Li}$ antisites and interstitial Ta$_{\rm V}$:V$_{\rm Li}$ defect pairs in one and the same crystal system.
As a consequence, the set of LN, LT, and LNT turns out to be a unique sequence of model systems for small polaron studies in systems with one (LN), two (LT) and three (LNT) intrinsic defect centers for electron localization, thus enabling the direct comparison to the formation, transport and recombination properties of the individual types of small bound polarons, but also for the study of their mutual interplay. Pulse-induced absorption enables the necessary experimental access to show-up significant differences in the comparison of the three isoelectric model systems, which is the topic of the present study.

We utilize a two-photon pump to produce small bound electron polarons whose recombination is then monitored by their absorption’s temporal decay. The transients of all LNT samples show a pronounced red and near-infrared absorption that is primarily attributed to small polarons according to previous findings in LN and LT \cite{Krampf2021,Enomoto-app11}. We explore the transient absorption as a function of pump intensity and composition $x$ from 10$^{-6}$ to 10$^2$\,s for temperatures between 290 K and 430\,K. A simple band model considering one, two and three intrinsic defect centers is used to relate our findings in LNT to remarkable differences in the small polaron hopping properties in direct comparison with LN and LT. The analysis of the activation energies, of the lifetimes and starting amplitudes of the light-induced signals in conjuncture with the intrinsic defect structure yield a plausible explanation for the increase of lifetime, the composition dependence of the small polaron number densities and the relaxation shape of the transients.
Based on this model approach, the possibility to generate different type of bipolarons by thermal annealing, but also of O$^-$ hole polarons and self-trapped excitons localized at Nb-O as well Ta-O actahedra by optical means can be concluded in analogy to the state of knowledge in LN. Moreover, we sketch the possibility to tailor the photoelectrical transport by composition in respect of the the small-polaron based photogalvanic effect, as well as to tailor LNT for applications in the field of nonlinear photonics by doping. 

\section{Materials and Methods}

\subsection{Single crystal growth and characterization}
LNT single crystals were grown from LiNb$_{1-x}$Ta$_x$O$_3$ melts using the Czochralski method with induction heating at the Leibniz-Institut für Kristallzüchtung (Berlin) as described in Ref.~\cite{Bashir-fe23}. In addition, c-LN and c-LT wafers were purchased from Precision Micro-Optics Inc., so that compositions from $x = 0.0 - 1.0$ were available for the study of this paper (see table~\ref{tab:table1}). 
\begin{table}[ht]
    \caption{c-LN, c-LT and LNT single crystals used in this study with crystal number, average Ta atomic fraction $x$ ($0\leq x\leq1$) as determined from X-Ray fluorescence, thickness $d$, type of crystal cut, optical absorption edge E$_{\rm gap}$ and source/supplier. PMO: Precision Micro-Optics Inc.; IKZ: Institut für Kristallzüchtung, Berlin}
    \label{tab:table1}
    \centering
    \begin{tabular}{cccccc}
    \hline
         crystal&\multirow{2}{*}{$x$}&\multirow{2}{*}{$d$ [mm]}&\multirow{2}{*}{cut}&\multirow{2}{*}{E$_{\rm gap}$ [eV]}&\multirow{2}{*}{supplier}\\
         identifier&&&&&\\
         \hline\hline
        c-LN&0.00&0.20&z-cut&3.97&PMO\\
        c-LN&0.00&0.49&x-cut&3.97&PMO\\
        LNT-05&0.06&0.92&x-cut&3.87&IKZ\\
        LNT-132&0.45&0.74&x-cut&3.97&IKZ\\
        LNT-01&0.45&1.00&z-cut&3.90&IKZ\\
        LNT-125&0.70&0.40&x-cut&4.09&IKZ\\
        LNT-110&0.89&1.94&x-cut&4.01&IKZ\\
        c-LT&1.00&0.50
        &z-cut&4.64&PMO\\
        c-LT&1.00&0.49&x-cut&4.64&PMO\\
         \hline
    \end{tabular}
\end{table}
Taking into account the congruent compositions of c-LN and c-LT of 48.38 mol\% Li$_2$O~\cite{Jundt-jcg08} and 48.46 mol\% Li$_2$O~\cite{Kushibiki-ffc06}, respectively, we consider an average value of 48.4 mol\% for all compositions. 

\begin{figure}
    \includegraphics[width=\textwidth]{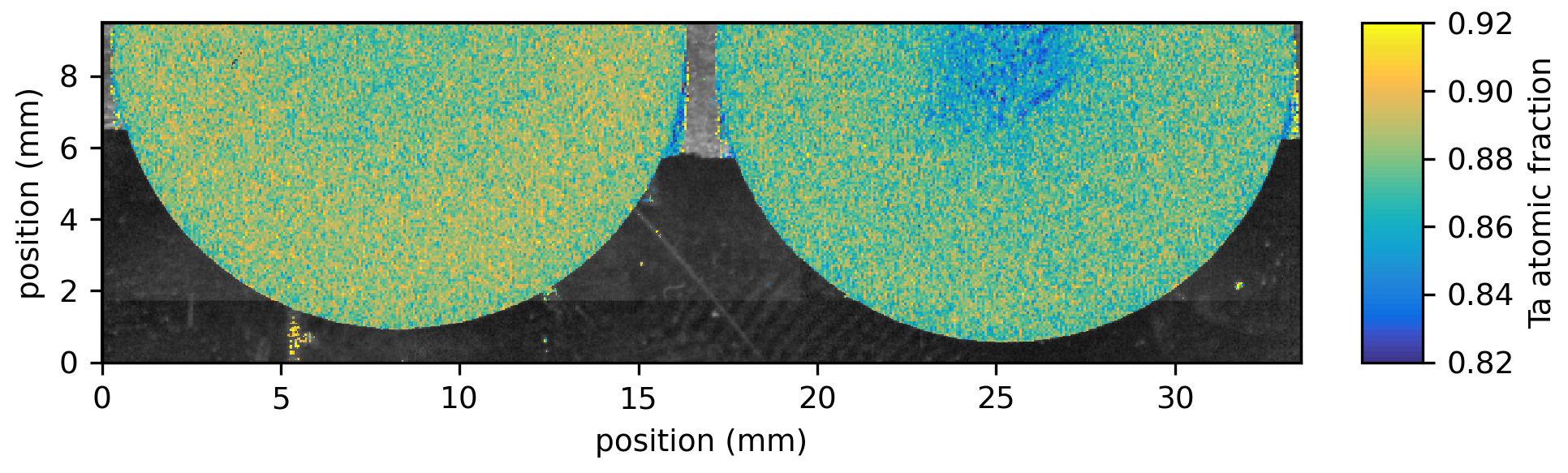}
    \caption{XRF elemental mapping of a LNT-110 plates ($x\approx0.89$) as cut from the as-grown crystal boule showing the Ta distribution. The local LT molar fraction $x$ of the LiNb$_{1-x}$Ta$_x$O$_3$ crystal is given as parameter aside the color scale.}
    \label{fig:figure1}
\end{figure}

Energy dispersive micro X-ray fluorescence ($\mu$-XRF) measurements were used to determine the elemental distribution and average concentration of Ta in the grown LNT single crystals as shown exemplarily for the LNT-110 sample with $x\approx0.89$ in Figure~\ref{fig:figure1}.
The measurements were carried out with a Bruker M4 TORNADO spectrometer at low vacuum environment (20 mbar). The X-ray source was a tube with Rh anode working at 50\,kV and 200$\mu$A. Bremsstrahlung was focused at the sample surface using polycapillary X-ray optics, yielding a high spatial resolution (beam width) of about 20\,$\mu$m with a measurement time per pixel of 20\,ms. The fluorescence signal was detected with a circular silicon drift detector. Quantification was done by the built-in routines of the spectrometer using a fundamental parameters database. The Ta concentration shows a gradual variation of a few percentage on the millimeter length scale with an increase of Ta from the rotation center to the rim of the boule. A similar gradient is observed along the growth direction as it can be concluded comparing the results for two cuts from the top and bottom of the boule (cut distance of about 10\,mm). In order to minimize the impact of a concentration gradient on our transient absorption measurements, the diameter of the probe beam ($<$ pump beam) is limited to less than 500\,$\mu$m (ns-studies) and 100\,$\mu$m (fs-studies) in all of our studies presented below. 

\begin{figure}[ht]
\centering
    \includegraphics{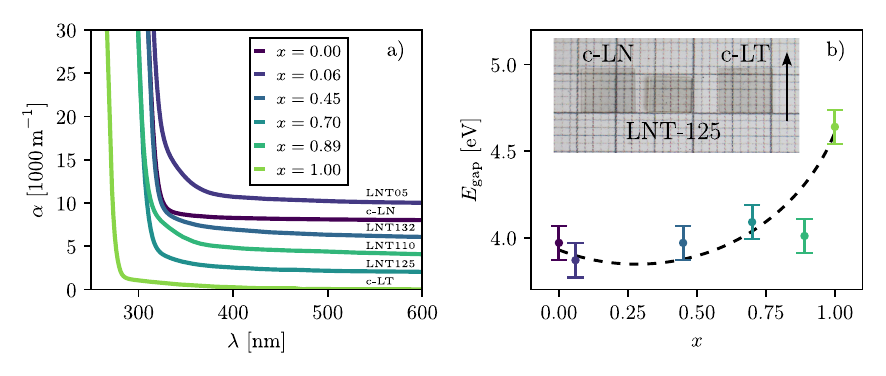}
    \caption{a) Absorption spectra of LN, LT, and LNT for various compositions over the visual spectral range. Spectra are stacked by $2000\,\textrm{m}^{-1}$. b) Optical band gap determined via Tauc-method for the spectra shown in a) Dashed line as guide to the eye.  The Inset shows a photograph of the a c-LN, c-LT and LNT-125 sample with the arrow indiating the orientation of the polar axis for all three samples.}
    \label{fig:figure2}
\end{figure}

For optical measurements, the crystal boule and wafers were cut into $x$- and $z$-plates with open aperture of $5\times 6$\,mm$^2$, ground to a thickness between $0.2-2.00$\,mm (cf. Table~\ref{tab:table1}) and finally polished to optical quality on both sides. The polar $c$-axis of the $x$-cut samples was aligned parallel to the shorter $6$\,mm side. Optical absorption spectra as a function of the Ta concentration were determined using a commercial two-beam spectrophotometer (Shimadzu Europa GmbH, UV-3600) in the ultraviolet-visible spectral region (see Figure~\ref{fig:figure2}a)).

All spectra show a broad-band transparency with values of the absorption coefficient $\alpha_0\ll 3$\,m$^{-1}$ over the entirely investigated spectral range. There is no evidence of the existence of impurities with relevant concentrations. In particular, we can neglect the absorption coefficients at the probing light wavelengths of $\lambda = 633$\,nm and $\lambda = 785$\,nm of the transient absorption experiments because E$_{\rm gap}$ is found at wavelengths below 320\,nm  (see Figure~\ref{fig:figure2}b)). The corresponding values were determined from the spectra using the Tauc-plot method and the procedure described in detail in Ref.~\cite{bock-applsci19,Tauc-pss66,Bhatt-pss12}. Here, an indirect, allowed transition is assumed~\cite{Wooten-book72}. As a result, E$_{\rm gap}$ of LN and LT as well as their mutual energetic difference are found in full accordance with the literature data ($\mathrm{E_{\rm gap}{\rm(LN)}=3.78}$\,eV \cite{Dhar-jap90, Thierfelder-pss10, zanatta-rip22} and $\mathrm{E_{\rm gap}{\rm(LT)}=4.5 - 4.7}$\,eV \cite{he-oc08, kato-cpl98, chao-apl96}). Moreover, the values of E$_{\rm gap}$ (cf. fig.~\ref{fig:figure2}b)) for LNT as a function of $x$ agree with the ones of Refs.~\cite{ pfannstiel-jpcm24, Klenen-nm24, Bernhardt-prm24}. 

\subsubsection{Nano-/femtosecond-pump continuous-wave probe spectroscopy}
Transient red and near-infrared absorption was inspected upon nanosecond and femtosecond pulse exposure using the two individual setups described in the following: For ns-pulse excitation, small polarons were generated in LNT via two-photon absorption using short intense laser pulses of an ordinarily polarized, frequency-doubled Nd:YAG (SpitLight 600, InnoLas Laser GmbH) pulse laser ($\lambda_{\rm p} = 532$\,nm, $\tau_{\rm FWHM}= 8$\,ns, $E_{\rm P}^{\rm max} \approx 400$\,mJ) according to the setups used for the inspection of LN in Refs.~\cite{Messerschmidt-ome19,Vittadello-cry18}. Transmitted polarized light of a cw diode laser ($\lambda= 785$\,nm) and a Helium-Neon laser ($\lambda=633$\,nm) was split and detected by a set of Si-PIN photo diodes, connected to a fast digital storage oscilloscope. Thus, the TA signal was recorded in a time range from nanoseconds up to five seconds and from one second up to several seconds, respectively. The power of the probe laser was kept below $1\,$mW to minimize its effect on the transient absorption. The transient absorption $\alpha_{\rm li}(\lambda, t)$ is determined from the ratio of the intensity of the transmitted probe light after the pulse was applied $I(\lambda, t)$ to that prior to the pulse event $I(\lambda, t \leq 0)$, giving: $\alpha_{\rm li}(\lambda, t) = -(1/d)\ln(I(\lambda, t)/I(\lambda, t \leq 0))$.

For the fs-pulse excitation a regeneratively amplified Ti:Al$_2$O$_3$ oscillator (Astrella, Coherent Corp.) is used to generate fs-pulses which are frequency doubled in a BBO crystal ($\lambda_{\rm P}=400$\,nm, $\tau_{\rm FWHM} \approx 60 \, \textrm{fs}$, $E_{\rm p} \approx 400\mu$J). The repetition rate of the laser is reduced by a pulse-picker to 10\,Hz. 
A further decrease of repetition rate is achieved by integration of a fast optical shutter into the pump beam path, enabling single pulse picking for measurements with relaxation times of well over 1 second. For probing, light of a focused Helium-Neon laser ($\lambda=633$\,nm) is superimposed with the pump pulse within the volume of the sample. The transmitted laser beam is focused via a lens onto a Si-PIN photo diode, measuring the transmitted intensity with a digital storage oscilloscope, which is triggered by residual pulse light, in two time ranges successively. In both setups, the samples are mounted onto a double-stack Peltier element, enabling temperature regulation in the range of 290 - 410\,K.

\section{Results}
\begin{figure}[ht]
    \centering
    \includegraphics{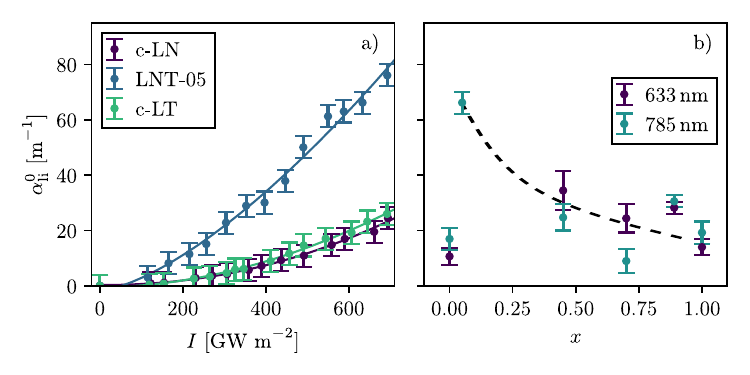}
    \caption{a) Initial absorption $\alpha_{\rm li}^0$ as a function of pump intensity ($\lambda_\mathrm{P}=532\,$nm, $\tau_\mathrm{P}=8$\,ns) for c-LN, c-LT and LNT-05 (representative for LNT) with a probing wavelength of 785\,nm,  fitted with eq. \ref{eq:equation1}. Resulting parameters are found in tab. \ref{tab:table2}. b) Composition dependent initial absorption $\alpha_{\rm li,0}$ for probing wavelengths of 785\,nm and 633\,nm. Excitation with the same laser system as in a), but at a fixed intensity of $I_\mathrm{P}\approx(600 \pm 50)$\,GWm$^{-2}$.
}
    \label{fig:figure3}
\end{figure}

In the first step, the generation of small polaron densities via multi-photon absorption process is validated by inspection of the starting amplitude of the light-induced absorption $\alpha_{\rm li}^0$ ($I_{\rm p}$) at 1\,$\mu$s after the pulse event  as a function of the pump pulse intensity in the range of $0<I_{\rm p}<7\cdot10^{11}$\,W/m$^2$ (see Figure\ref{fig:figure3}a)).
The results are plotted for excitation with ns-laser pulses at a  wavelength of $\lambda=532$\,nm (extraordinary light polarization) and a probing wavelength of $\lambda=785$\,nm showing a nonlinear growth of the induced absorption with increasing intensity (repetition rate 0.1\,Hz). For comparison, the results for c-LN and c-LT are plotted, as well. 
The solid line represents the result of a fit of the function 
\begin{equation}\label{eq:equation1}
\alpha_{\rm li,0} = m\cdot I^{\kappa}+b
\end{equation}
to the experimental data set with the fitting parameters for each sample and probing wavelength given in tab~\ref{tab:table2}. Both c-LN and c-LT samples are well described by a quadratic intensity dependence. However, for the LNT-05 sample, at elevated intensities a deviation from the quadratic dependence is noticeable, that may point to a saturation behavior as reported for LN in Ref.~\cite{Merschjann-jpcm09}. Overall the exponent for this sample is found to be around $\kappa\approx1.5$ although with a much higher prefactor of $m\approx5.2\cdot10^{-3}\,\textrm{m}^{2}\textrm{GW}^{-1.5}$ compared to c-LN and c-LT which feature values in the range of $\sim 10^{-5}\,\textrm{m}^{3}\textrm{GW}^{-2}$.

\begin{table}[ht]
\caption{Fitting results of the prefactor $m$, exponent $\kappa$ and offset $b$ of eq.~\ref{eq:equation1} to the intensity dependent initial absorption $\alpha_{\rm li}^0$ for the curves shown in fig.~\ref{fig:figure3}a).}
\label{tab:table2}
\centering
\begin{tabular}{cccc}
\hline
\multicolumn{1}{l}{sample} &  $m$ [m$^{2\kappa-1}$GW$^{-\kappa}$] & $\kappa$ & $b$ [m$^{-1}$] \\
\hline\hline
\multirow{1}{*}{c-LN}&$(2.2\pm1.7)\cdot10^{-5}$&$2.1\pm0.1$&$\phantom{\mathrm{-}}0.3\pm0.4$\\
LNT-05               &$(5.2\pm4.1)\cdot10^{-3}$&$1.5\pm0.1$&$-1.8\pm2.3$\\
\multirow{1}{*}{c-LT}&$(9.2\pm5.5)\cdot10^{-5}$&$1.9\pm0.1$&$-0.5\pm0.4$\\
\hline
\end{tabular}
\end{table}
Figure~\ref{fig:figure3}b) shows the starting amplitude of the light-induced absorption as a function of composition for probing with light in the red ($\lambda=633$\,nm) and near-infrared ($\lambda=785$\,nm) spectral range.
Both probing wavelengths show a similar composition dependence with pronounced non-linearity. The starting amplitude jumps to a 7x fold larger value at $x=0.05$ and decreases with increasing Ta-content to the end level of c-LT, that itself corresponds to the one of LN.
      \begin{figure}[ht]
        \centering
        \includegraphics{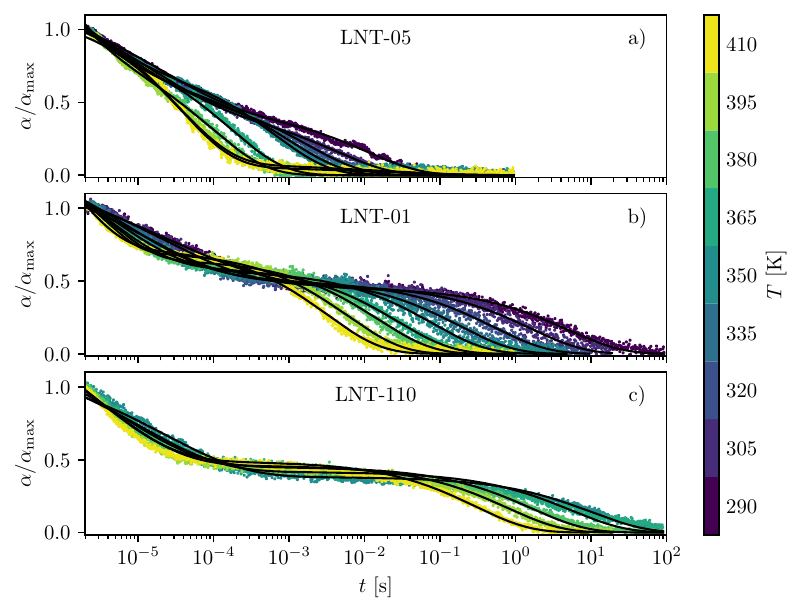}
        \caption{Transients of a) LNT-05 b) LNT-01 and c) LNT-110 at various temperatures excited via fs-setup ($\lambda_\mathrm{P}=400$\,nm, $\tau_\mathrm{P}\approx60$\,fs, $I_\mathrm{P}\approx5$\,PWm$^{-2}$) with a probing wavelength of $\lambda=633$\,nm.}
        \label{fig:figure4}
    \end{figure} 

\begin{figure}[ht]
    \centering
        \includegraphics{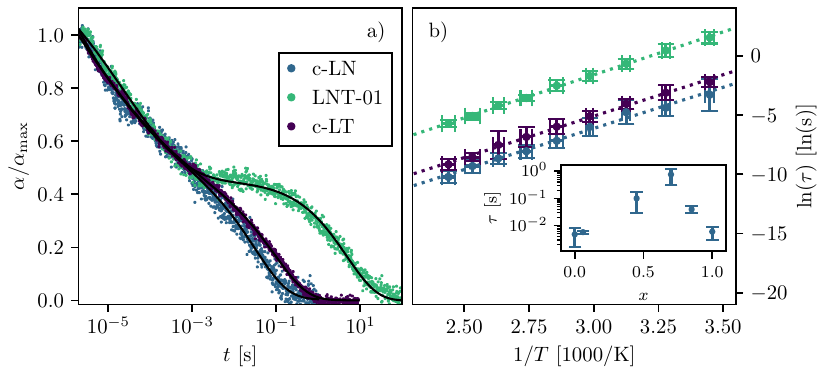}
        \caption{a) Transients of c-LN, c-LT and LNT-01 at $T=290$\,K via fs-excitation ($\lambda_\mathrm{P}=400$\,nm, $\tau_\mathrm{FWHM}\approx60$\,fs, $I_\mathrm{P}\approx5$\,PWm$^{-2}$) with a probing wavelength of $\lambda=633$\,nm.
        b) Lifetime $\tau$ vs temperature $T$ in an Arrhenius plot for c-LN, c-LT and LNT-01. The inset shows the 1/e-lifetimes of the transient absorption signal of various samples.}
        \label{fig:figure5}
    \end{figure}

Figure \ref{fig:figure4} shows the transients of LNT-01, LNT-05 and LNT-110 for different temperatures recorded with the fs-setup. A two-stage decay can be observed whereby the decay times decrease with increasing temperature. With increasing Ta content, the overall relaxation time rises disproportionally. The sum of two independent Kohlrausch-Williams-Watts (KWW) functions:
\begin{equation}\label{eq:equation2}
    \alpha_{\rm li}(t)=\alpha_0\exp[-(t/\tau)^\beta]
\end{equation}
with the lifetime $\tau$ and the stretching exponent $\beta$ is fitted to the data set.
For direct comparison, the LNT-01 transient of 290\,K is shown in fig.~\ref{fig:figure5}a) together with the respective transient absorption curves for c-LN and c-LT. While the initial shape of the decay curves superposes well between the three samples, an additional shoulder develops in the c-LT and LNT-01 transients at larger probing times. In c-LT, this shoulder is of minor amplitude and modifies the decay process only slightly. The corresponding shoulder in the LNT-01 sample is much more expressed and causes an elongation of the transients of multiple orders of magnitude. The solid lines in fig.~\ref{fig:figure5}a) represent fits of a stretched exponential decay to the datasets. While the c-LN and c-LT transients could sufficiently be described by a single KWW-fit, the LNT-01 measurement can only be reproduced by a sum of two such functions where one corresponds to an initial, fast decay and one corresponds to a second, slow decay. The temperature dependence of the determined lifetimes are depicted in fig.~\ref{fig:figure5}b) in an Arrhenius plot. Here, for the LNT-01 sample, the fitted lifetime of the slow, second decay is shown. The temperature dependence is modeled by
\begin{equation}\label{eq:equation3} 
    \tau(T) = \frac{1}{Z} \exp{ \left(\frac{E_a}{k_{\rm B} T} \right)}
\end{equation}
with the frequency factor $Z$ and the Activation Energy $E_{\rm a}$. The data adheres to this Arrhenius behaviour well within the scope of the respective uncertainties. The obtained activation energies are $(0.60 \pm 0.05)\, \textrm{eV}$ for c-LN and c-LT and $(0.63 \pm 0.05)\,\textrm{eV}$ for LNT-01 so that they do not show a significant difference. The inset of fig~\ref{fig:figure5}b) shows the compositional dependence of the measured 1/e-lifetimes of all investigated samples after ns-excitation at room temperature. The lifetimes spread over multiple orders of magnitude with the value established for c-LN forming the lower boundary. The c-LT sample features a marginally increased lifetime compared to c-LN. The peak lifetime is found for the LNT-110 sample of $x\approx0.85$. From this peak value the lifetimes scale exponentially toward the edge compositions of c-LN and c-LT. 

\section{Discussion}
Our experimental study shows that nano- and femtosecond pulse-induced red and near-infrared transient absorption is present in LNT crystals for all compositions $x$ inbetween the edge systems LN and LT. In particular, we find that the effect shows a number of similarities to the phenomena of green-light, blue-light or ultraviolet-light induced infrared absorption (GRIIRA, BLIIRA, UVIIRA) that has been widely studied in LN and LT \cite{Furukawa-ap01, Nakamura-fe02,Ali-ass11,Hirohashi-aip07}, so far. This includes a nonlinear intensity dependence of the starting amplitude $\alpha_{\rm li}^0$, an induced absorption at photon energies of 1.6\,eV and 2.0\,eV, a stretched-exponential decay of the transients with two distinct decay channels, a lifetime exceeding the millisecond time range, and an activation energy in the order of $E_\mathrm{a}\approx0.6$\,eV. Beyond that we find that the absolute values of the 1/e-lifetimes of LNT exceed the ones of LN and LT by up to two orders of magnitude (cf. inset of fig~\ref{fig:figure5}b)), thus pointing to a strongly delayed decay process. As a particular feature of LNT, the band gap energy, as well as the starting amplitude $\alpha_{\rm li}^0$ show a characteristic $x$-dependence with the minimum of the band gap and the maxima of the starting amplitude both experimentally determined at $x=0.06$.

Taking all these findings for LNT and the state-of-the-art knowledge for LN and LT into account, it is reasonable to assume that the pulse-induced absorption feature is fundamentally of the same physical origin in all three model systems: the formation of optically generated small bound polarons on the sub-ps time scale~\cite{Krampf2021,Krenz2022}, the hopping transport of small bound polarons on timescales up to milliseconds~\cite{Vittadello-cry18,Messerschmidt-ome19} and the recombination with holes that terminates the induced absorption~\cite{Messerschmidt-ome19, Schirmer2006}. In what follows we will therefore sketch a possible scenario that is capable to explain the findings in LNT by optically induced small bound electron polarons in a straightforward and comprehensive manner, but also to enable a first insight to the discovered significant differences between the three isostructural ferroelectrics. For this purpose, differences in the intrinsic defect structures are considered and turn-out to be sufficient to reconstruct all our findings. 

\begin{figure}[ht]
        \includegraphics[width=0.99\textwidth]{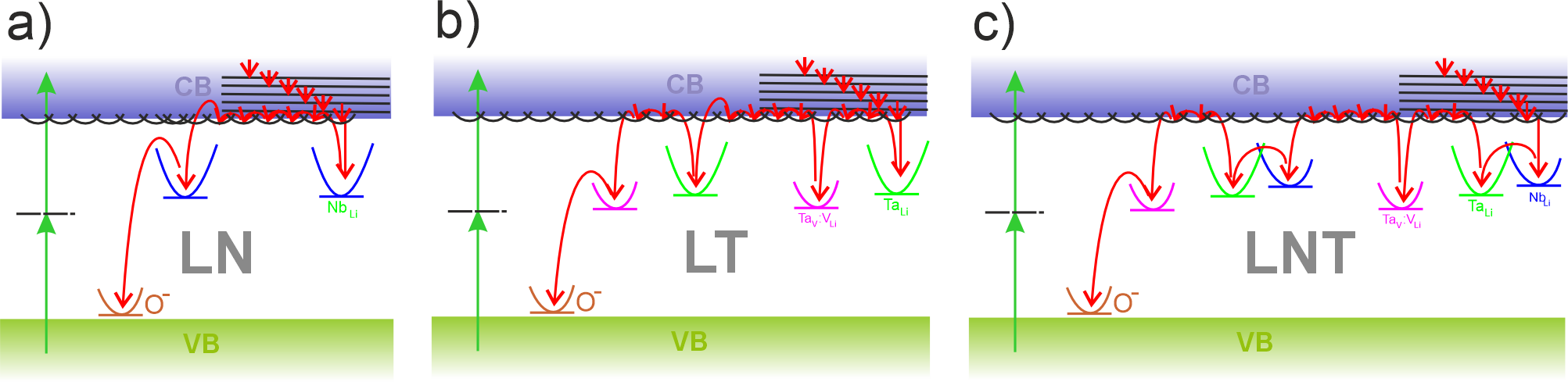}
        \caption{a) Schematic plot of excitation, transport and relaxation scheme for LN. b) Schematic plot of excitation, transport and relaxation scheme for LT. c) Schematic plot of excitation, transport and relaxation scheme for LNT}
        \label{fig:figure6}
    \end{figure}

The anchor of our model approach is to consider the unique situation of three intrinsic defect centers for electron self-localization in the model system LNT: the Nb$_{\rm Li}$ and Ta$_{\rm Li}$ antisite as-well-as interstitial Ta$_{\rm V}$:V$_{\rm Li}$ defects. This distinguishes LNT from LN and LT which show only one (Nb$_{\rm Li}$ antisite in LN) and two (Ta$_{\rm Li}$ antisite and Ta$_{\rm V}$:V$_{\rm Li}$ interstitial defect in LT) intrinsic defects. Following this approach, the process of optical excitation, small polaron transport and recombination must be different for LNT in comparison to LN and LT as schematically sketched in the band models of Figure~\ref{fig:figure6}. 

\noindent
\textit{Lithium niobate}\\
\noindent
For simplicity, we will first recall the situation for c-LN with only one defect center. According to the state-of-the-art knowledge, the absorption features of Nb$_{\rm Li}^{4+}$ antisite polarons show the following characteristics~\cite{Merschjann-jpcm09,Schirmer2009}: an optical absorption band centered at $\sim1.6$\,eV with a width of $\sim1.0$\,eV, and an absorption cross section of $\sigma_{\rm pol}=(4-14)\times10^{-22}$\,m$^2$. This enables to probe the polaron state at $\sim1.6$\,eV with sufficient signal-to-noise ratio (cf. Fig.~\ref{fig:figure3}). For LN, grown from the congruent melt, we can
estimate the number density of antisites to $n_{\rm antisites}\approx 19\cdot 10^{25}$\,m$^{-3}$, that is an upper limit for the possible number density of pulse-induced small polarons $n_{\rm pol}$ and related to the starting amplitude in Fig.~\ref{fig:figure3} via $\alpha_{\rm li}^0=n_{\rm pol}\cdot \sigma_{\rm pol}$. On the sub-100\,fs time scale, 
exposure to single (ultra-)short, intense laser pulses results in the generation of a large number density of hot carriers in the conduction band via two-photon absorption, that is reflected by the quadratic increase of $\alpha_{\rm li}^0$ as a function of intensity (cf. fig.~\ref{fig:figure3}a)), in good agreement with literature \cite{Freytag2018}. Subsequently cooling to the conduction band minimum, followed by self-trapping events which can result in the formation of short-lived excitonic states or small polarons either at regular Nb$_{\rm Nb}$-sites (small free Nb$_{\rm Nb}^{4+}$ electron polaron) or at Nb$_{\rm Li}$-antisites (small bound Nb$_{\rm Li}^{4+}$ electron polaron).
Accordingly, holes are formed for charge compensation that potentially result in the formation of small O$^-$ hole polarons~\cite{Schirmer2006}. The latter show characteristic absorption features with maxima in the blue-green spectral range~\cite{halliburton-prsb84} and, thus, are not considered in the understanding of the red and near-infrared transients. Transport of the bound electron polarons occurs via 3D hopping through the regular and intrinsic defect landscape so that intermediate trapping at antisites occurs~\cite{Emin2013}. Experimentally, a fingerprint for hopping transport in the non-regular LN lattice is the appearance of a non-exponential decay behavior of the transient absorption as a result of distant-dependent excitation and recombination 
rates~\cite{berben-jap00}. The stretched exponential absorption kinetics is best reconstructed using a Kohlrausch-Williams-Watt-function (cf. eq. ~\ref{eq:equation2} ), with the stretching exponent $0\leq \beta\leq 1$ and $\beta\approx 0.6$ for c-LN. Non-radiative electron-hole recombination occurs if small bound polarons and holes are in direct vicinity to each other. Transient absorption terminates with the complete death of small bound polarons.\\\\
\noindent
\textit{Lithium tantalate}\\
\noindent
The situation in LT is similar, but a second defect center, the Ta$_{\rm V}$:V$_{\rm Li}$ defect pair, must be considered~\cite{Vyalikh-prm18} for intermediate polaron trapping within the formation, 3D hopping transport and recombination path. The bound polaron state at the interstitial defect, that has been theoretically modeled and experimentally validated in Ref.~\cite{pfannstiel-jpcm24}, shows a maximum of its optical absorption band at $\sim1.5$\,eV, a width of the optical absorption band of $\sim1.0$\,eV, while the Ta$_{\rm Li}^{4+}$ antisite polaron is blue-shifted by about 0.5\,eV. The latter is a result of the comparably larger band gap energy of c-LT (4.8\,eV \cite{kato-cpl98,he-oc08}) in comparison with c-LN (4.1\,eV \cite{Bhatt-pss12,zanatta-rip22}). Therefore, a two-photon excitation of small bound polarons is expected for LT, as well, and experimentally validated by our experimental findings in Fig.~\ref{fig:figure3}. We note, that fs-pulse excitation can be used, as well (cf. Ref.~\cite{Krampf2021}) and reveals similar kinetics in the small polaron formation scheme as LN. Here, the upper bound of pulse-incuded small polaron densities in LT can be estimated to $n_\mathrm{defects}\approx 15\cdot 10^{25}$\,m$^{-3}$ according to the number density of intrinsic defects calculated from the excess of Ta ions in LT grown from the congruent melt. So far, there is no report on the absorption cross section for both of the electron polaron states in LT. However, first of all it seems reasonable to assume that it is comparable to the one of LN~\cite{Merschjann-jpcm09}. We note, however, that the formation of antisite polarons is slightly preferred in comparison to the interstitial polaron state~\cite{pfannstiel-jpcm24,Vyalikh-prm18}. With respect to the transient absorption on the long-term, a stretched exponential decay behavior is experimentally found, again, (cf. Fig.~\ref{fig:figure5}) pointing to the presence of small polaron 3D hopping transport in LT, as well. Sub-ns transients were studied in Ref.~\cite{Krampf2021} and showed a temporal constant signal up to a delay of 4\,ns with respect to the fs-pump.
This behavior is validated with the resulting lifetimes of the KWW fits, shown in fig. ~\ref{fig:figure5}. For c-LT a $\tau$-value of $\sim 108$\,ms is found, about a factor of three larger compared to c-LN. From the viewpoint of intrinsic defects, this change goes along with the presence of an additional trapping center. The slightly lower value of the total electronic polaron energy of the interstitial defect with a difference of $\sim 0.1$\,eV to c-LN, as well as a blue-shift of the Ta$_{\rm Li}$ antisite are both factors that affect the hopping transport mechanism and, thus, are capable to explain the increased lifetime of the transient.\\\\

\noindent
\textit{Lithium niobate tantalate}\\
\noindent
Let us now turn to the model system LNT with the assumption of three bound polaron states in its defect landscape as depicted schematically in Fig.~\ref{fig:figure6}c). We consider, that the electronic (polaron binding energy) and optical features (position of the absorption maxima, width of the absorption feature, absorption cross section) of the respective electron polarons largely correspond to those of c-LN and c-LT. The increased lifetime and characteristic shape of the LNT transient absorption correspond with the three defect centers as follows:

The stretched exponential decay as well as the activation energy of $\sim0.6$\,eV found in the Arrhenius like behaviour for all three depicted samples are strong indications that all polarons involved are bound states of similar activation barrier, which dominate the hopping transport. The elongated lifetime is explained by the vertical shift in the Arrhenius behaviour (cf. fig. \ref{fig:figure5}b)) resulting in a reduced frequency factor in LNT compared to c-LN and c-LT and therefore a decreased effective jump rate, slowing the polaron transport. According to the Marcus-Holstein model this frequency factor is determined via $Z\propto\exp(-d)$, where $d$ is the radial distance between the sites participating in a polaron jump event \cite{Austin-adp01}. In case of two or more trapping centers, the transport is dominated by the features of the center with the lowest total energy and/or largest activation energy. In the present case, this is the antisite Ta$_{\rm Li}$ state~\cite{pfannstiel-jpcm24}. Since the overall defect density is approximately of the same order in c-LN and c-LT, the density of Ta$_{\rm Li}$ sites in c-LT has to be lowered considerably compared to the Nb$_{\rm Li}$ antisite density in c-LN. The average radial distance $d$ between Ta$_{\rm Li}$ in c-LT is therefore increased compared to the Nb$_{\rm Li}$-Nb$_{\rm Li}$ distance in c-LN, slowing down the transport and elongating the bound polaron lifetimes through a reduction of $Z$. This is already seen in the sub-ns time scale where the decay mechanism of the near-infrared transients is considerably decelerated in c-LT compared to c-LN~\cite{Krampf2021}. Under consideration of a similar total defect concentration in LNT, as suggested in literature \cite{bartasyte-MatChemPhys12}, this effect is pronounced even stronger since the number density of Ta$_{\rm Li}$ defect sites shrinks in order to accommodate a total of three intrinsic defects in the lattice. The clear dependence of the total lifetime of the transients in function of the Ta atomic fraction (cf fig.~\ref{fig:figure4} and fig.~\ref{fig:figure5}) suggests that the internal defect structure, in particular the ratio between the three different atomic defects, may change substantially over the compositional range.
The presence of an additional defect state, i.e. the interstitial Ta$_{\rm V}$:V$_{\rm Li}$ defect, remains responsible for the lifetime extension.

This consideration also allows for an understanding of the increased starting amplitude $\alpha_{\rm li}^0$ and of the non-linear exponent $\kappa$. In particular it is found that the prefactor $m$ of LNT is substantially increased compared to c-LN and c-LT for all compositions. This means that the number density of polarons at the beginning of the measurement regime $t\approx10^{-6}$\,s is increased significantly in LNT as less polarons recombine during the unprobed time of $t<10^{-6}$\,s after excitation. With respect to $\kappa$, there is no doubt that the intensity dependence of the starting amplitude $\alpha_{\rm li}(I)$ of Fig.~\ref{fig:figure3}a) is non-linear, however, the resulting exponent $\kappa=1.5$ does not show a quadratic dependence as it does for c-LN and c-LT. This difference is due to the methodological influence of the particularly long lifetimes of the transient absorption on consecutive pulses in our measurement protocol. Accordingly, it can be assumed that a two-photon process is also present for LNT and the main contribution to the excitation.

The following conclusive explanation for pulse-induced transients in LNT emerges from the aforementioned aspects: (i) small bound electron Nb$_{\rm Li}$, Ta$_{\rm Li}$ antisite polarons and Ta$_{\rm V}$:V$_{\rm Li}$ interstitial polarons are optically generated simultaneously and without special preference with single fs- or ns-pulses, (ii) all three polaron states act as intermediate trapping centers in the thermally activated 3D hopping transport, while the presence of Ta$_{\rm Li}$ antisite defect shows the largest impact on slowing-down the process, (iii) the control of the transients by composition and defect tuning is a direct result of the possibility to tailor the relative concentration of three intrinsic defects in LNT and reaches extreme cases at $x=0$ and $x=1$ due to the elimination of one (c-LT) and two (c-LN) defect centers. Consequently, it can be stated that the (macroscopic) photoelectric properties of LNT are determined by the combination of small polarons and intrinsic defect structure to a large extent. This in turn can be a starting point, e.g., for the investigation of the volume photovoltaic effect in LNT with the aim of further developing the associated microscopic modelling based on small polarons\cite{Schirmer2011}. In addition, it can be expected that two charge carriers can also be localised as bipolarons at the aforementioned defect centres and thus up to three bipolaronic states can occur simultaneously, e.g. in LNT after thermal annealing. Furthermore, hole polarons are to be expected, so that overall a supplementary study on pulse-induced absorption in the blue-green spectral range is promising. \\\\

\section{Summary \& conclusion}

We can conclude that the presence of three ($\mathrm{Nb_{Li}}$,  $\mathrm{Ta_{Li}}$ antisites and $\mathrm{Ta_{V}}$:$\mathrm{V_{Li}}$ interstitial) intrinsic defects in LNT in contrast to one ($\mathrm{Nb_{Li}}$ antisite in LN) and two ($\mathrm{Ta_{Li}}$ antisite and $\mathrm{Ta_{V}}$:$\mathrm{V_{Li}}$ interstitial in LT) result in major changes to the small polaron thermally activated 3D hopping transport. From these three intrinsic defect centers, $\mathrm{Ta_{Li}}$ has the highest thermal activation barrier and thus dominates the polaron transport. Its number density can be correlated reciprocally to the decay time of the transients and therefore to the polaron lifetime. The latter varying by two orders of magnitude over the enitre compositional range. As such, the binary system LN-LT is identified as a unique platform for the engineering of (photoelectrical) charge transport properties, well beyond the capabilities of its edge compounds LN and LT, via defect and stoichiometry adjustments. This enables a refinement of the impact of optoelectrical phenomena like stronger charge separation in case of the photorefractive and bulk photovoltaic effects or higher current densities for the small-polaron based photovoltaic effect~\cite{Schirmer2011}. Retroactively, the findings for LNT may be applied to LN and particularly LT in the same manner, expanding the parameter space for the engineering and application tuning of these well understood materials.
\section{Acknowledgment}
We gratefully acknowledge financial support by the Deutsche Forschungsgemeinschaft (DFG) through the research group FOR5044 (Grant No. 426703838, projects IM37/12-1, GA2403/7-1).

\section*{References}
\bibliographystyle{iopart-num}
\bibliography{manuscript.bib}

\end{document}